\documentclass[aip,jcp,twocolumn,10pt,superscriptaddress,showpacs,floats,floatfix,preprintnumbers,amsmath,amssymb]{revtex4-1}
\usepackage{graphicx}
\usepackage{dcolumn}
\usepackage{bm}

\graphicspath{{./fig/}}
\newcommand{\figwidth}{8.0cm}
\newcommand{\dd}{{\rm d}}
\renewcommand{\r}{{\bf r}}

\begin{document}

\author{Alex Willand}
\affiliation{Department of Physics, Universit\"{a}t Basel, Klingelbergstr. 82, 4056 Basel, Switzerland}

\author{Yaroslav O. Kvashnin}
\affiliation{European Synchrotron Radiation Facility, 6 Rue Jules Horowitz, BP220, 38043 Grenoble Cedex, France}

\author{Luigi Genovese}
\affiliation{Laboratoire de simulation atomistique (L\_Sim), SP2M, UMR-E CEA / UJF-Grenoble 1, INAC, F-38054 Grenoble,
France}

\author{\'Alvaro V\'azquez-Mayagoitia}
\affiliation{Argonne Leadership Computing Facility, Argonne National Laboratory,
IL, USA, 60439}

\author{Arpan Krishna Deb}
\affiliation{Laboratoire de simulation atomistique (L\_Sim), SP2M, UMR-E CEA / UJF-Grenoble 1, INAC, F-38054 Grenoble,
France}

\author{Ali Sadeghi}
\affiliation{Department of Physics, Universit\"{a}t Basel, Klingelbergstr. 82, 4056 Basel, Switzerland}

\author{Thierry Deutsch}
\affiliation{Laboratoire de simulation atomistique (L\_Sim), SP2M, UMR-E CEA / UJF-Grenoble 1, INAC, F-38054 Grenoble,
France}

\author{Stefan Goedecker}
\affiliation{Department of Physics, Universit\"{a}t Basel, Klingelbergstr. 82, 4056 Basel, Switzerland}

\title{Norm-conserving pseudopotentials with chemical accuracy compared to all-electron calculations}

\begin{abstract} 
By adding a non-linear core correction to the well established Dual Space Gaussian type pseudopotentials for the chemical elements up to the third period, we construct improved pseudopotentials for the Perdew Burke Ernzerhof (PBE) [J. Perdew, K. Burke, and M. Ernzerhof, Phys. Rev. Lett. 77, 3865 (1996)] functional and demonstrate that they exhibit excellent accuracy. 
Our benchmarks for the G2-1 test set show average atomization energy errors of only half a kcal/mol. The pseudopotentials also remain highly reliable for high pressure phases of crystalline solids.
When supplemented by empirical dispersion corrections [S. Grimme, J. Comput. Chem. {\bf 27}, 1787 (2006); S. Grimme, J. Antony, S. Ehrlich, and H. Krieg, J. Chem. Phys. {\bf 132}, 154104 (2010)] the average error in the interaction
energy between molecules is also about half a kcal/mol.
The accuracy that can be obtained by these pseudopotentials in combination with a systematic basis set is well superior 
to the accuracy that can be obtained by commonly used medium size Gaussian basis sets in all-electron calculations. 
\end{abstract}

\maketitle

\section{Introduction}

During the last decades, density functional theory (DFT) has proven its pivotal role for computational studies in the fields of condensed matter physics and quantum chemistry. Particularly the Kohn-Sham (KS) formalism of DFT has gained enormous popularity as an \textit{ab initio} method applicable to relatively large systems. An essential ingredient for many large scale implementations of KS-DFT are pseudopotentials which are also frequently denoted as effective core potentials. 
By eliminating the strongly bound core electrons pseudopotentials reduce the number of occupied electronic orbitals that have to be treated in an electronic structure calculation. In addition the valence wavefunctions arising from a pseudopotential are much smoother than the all-electron valence wavefunction in the core region, since the orthogonality constraints to the rapidly-varying wavefunctions carrying core electrons are missing. 
Since it is not necessary to describe rapidly varying wavefunctions the size of the basis set used for their representation can be reduced.
These two factors lead to a significant reduction of the computational effort of a pseudopotential calculation compared to an all-electron calculation. 

Even though it is well known that the valence electrons are responsible for the majority of chemical and physical
properties of atoms, pseudopotentials have to be constructed very carefully in order to reproduce the properties of the
 all-electron atom accurately. If a pseudoatom, i.e an atom described by a pseudopotential, reproduces the all-electron
behavior accurately for any chemical environment the pseudopotential is said to be transferable. 

Pseudopotentials (PSP's) are an essential ingredient of most electronic structure codes and different solutions are implemented in present-day DFT codes. Traditional norm-conserving (NC) approaches, e.g.~\cite{Chiang} are formally the simplest approach, since they give rise to pseudowavefunctions which lead to a valid charge density. By introducing atomic like orbitals as additional basis functions any atomic Hamiltonian arising either from an all-electron potential or from a norm conserving pseudopotential~\cite{lippert} can be transformed into an Linearized Augmented Plane Wave (LAPW) like Hamiltonian~\cite{singh,maschke,mylapw}. The widespread Projector-Augmented Wave (PAW) methods~\cite{PAW} and the ultrasoft pseudopotentials~\cite{Vanderbilt} are derived by such a transformation from an all-electron atom Hamiltonian. The number of required basis functions is reduced by this transformation, but the calculation of the charge density is more complicated and a generalized eigenvalue problem has to be solved even for the case of an orthogonal basis set.
For applications in quantum chemistry, effective core potentials~\cite{Stuttgart-ECP,SBKJC-ECP,LANL2DZ-ECP} are often optimized for a certain basis set and usually employed for heavier elements only.  

In this paper, the Dual Space Gaussian pseudopotentials of Goedecker-Teter-Hutter (GTH) and Hartwigsen-Goedecker-Hutter (HGH)~\cite{Hartwigsen98} PSP are generalized by the inclusion of a Non-Linear Core Correction (NLCC) term.
These new pseudopotentials are able to provide an accuracy that is comparable to that of the very best all-electron calculations.

The starting point for understanding why pseudopotentials work is the subdivision of space into muffin-tin spheres centered on the atom in a molecule or solid and the remaining interstitial region~\cite{martin}. A non-selfconsistent Schr\"{o}dinger equation can be solved exactly in the interstitial region if one knows the scattering properties on the surface of the muffin-tin spheres~\cite{heine}. The scattering properties are typically specified by the logarithmic derivative as a function of energy. 
This function is the quotient of the radial outward derivative and the functional value of the wavefunction on the
surface of the muffin-tin sphere. 
In this way the boundary conditions are specified which are necessary to integrate the Schr\"{o}dinger equation, a second order partial differential equation where the amplitude of the solution is fixed by a normalization constraint. 
A necessary but not sufficient condition for a pseudopotential to be transferable is therefore that the logarithmic derivatives of the all-electron and pseudo-atom agree over the relevant energy interval.  
The construction of pseudopotentials is typically done using as the reference state a neutral isolated atom which has been spherically symmetrized. This symmetrization can be achieved by using identical and generally fractional occupation numbers for all the orbitals with the same $n$ and $\ell$ quantum numbers, e.g. for the set of $2p_x$, $2p_y$ and $2p_z$ orbitals. 
The norm conservation condition~\cite{normcons} ensures that the logarithmic derivative function describes well the scattering properties of a muffin-tin sphere containing the charge distribution of this reference configuration. 
In a selfconsistent calculation the charge distribution changes however when the free atom is inserted in a molecule or solid and the potential in the muffin-tin region will in general differ from the potential within a muffin-tin sphere of the same radius around the reference atom. Hence the scattering properties change and the pseudopotential constructed using the charge distribution in the muffin-tin sphere of the isolated atom might not well reproduce these modified scattering properties of a new chemical environment. 
Due to screening effects there exists however an invariant muffin-tin sphere within which the total electronic charge distribution is nearly independent of the chemical environment~\cite{transferability}. 
The radius of this invariant muffin-tin sphere is a fraction of the covalent bondlength and thus considerably smaller than the muffin-tin radii used in other methods such as the LAPW method. 
The scattering properties of this invariant muffin-tin sphere hardly vary as a function of the chemical environment of the atom. 
If the separable terms of a pseudopotential as well as the difference between the local part of the pseudopotentials and the pure coulombic potential remain localized within this invariant sphere the pseudopotential is expected to be highly transferable. This recipe was followed in the construction of the GTH~\cite{Goedecker96} and HGH~\cite{Hartwigsen98} pseudopotentials which are indeed well transferable for non-spinpolarized systems. 

\begin{figure}
\includegraphics[angle=-90, width=\figwidth]{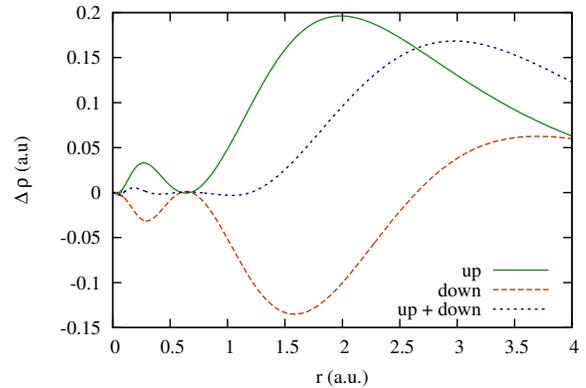}
\caption{ \label{Ptransfer} Difference in the radial spin densities and total charge density when adding half an electron to a phosphorus atom ($ [Ne] 3s^2 3p^{3.5} $). The inert core region of the total charge is not observed for the individual spin densities.}
\end{figure}

Despite the fact that the total charge in the invariant muffin-tin sphere is nearly identical in different chemical environments, the spin polarization is not, as illustrated in Fig.~\ref{Ptransfer}. Shown are the changes in the radial charge and spin densities if one adds half an electron to the unoccupied spin channel of the $3p$-orbital.
For spin polarized calculations the concept of an invariant muffin-tin sphere is therefore not applicable. One possibility to overcome this problem is to construct pseudopotentials which have an explicit spin dependence~\cite{carter}. 
The other possibility is to include nonlinear core corrections (NLCC)~\cite{Louie82} in the pseudopotential.

In the NLCC schemes the spin and charge densities in the muffin-tin sphere are not just the ones from the valence electrons treated explicitly by the pseudopotential but they are both the respective sums of the valence charge and the core charge given by the nonlinear core correction. Since the core electrons can be considered to be frozen, i.e to be invariant with respect to different chemical environments, this core charge is fixed once and for all. 
It is obvious that the spin polarization, i.e. the quantity
 $\theta({\bf r}) = \left(\rho_{up}({\bf r}) - \rho_{down}({\bf r})\right) /
                    \left(\rho_{up}({\bf r}) + \rho_{down}({\bf r})\right) $
is very poorly represented without a core charge. If for instance all valence orbitals are spin up then the spin down charge density $\rho_{down}({\bf r})$ would be zero and the spin polarization $\theta$ would be equal to one. In the real atom $\theta$ is not equal to one in the core region since the core electrons are never spin polarized. Since the density of the core electrons is much larger than the density of the valence electrons in the core region the spin polarization actually has typically small values. These correct small values of $\theta$ are reestablished by the core charge of a NLCC pseudopotential and exchange correlation functionals can provide reliable total energies. 
Nonlinear core corrections have therefore the potential to substantially improve the description of spin polarized states. 

Whereas previous implementations of NLCC pseudopotentials ~\cite{Porezag99} tried to faithfully represent the core charge, we follow here another approach. In the spirit of the GTH pseudopotentials where all terms have simple parametrized analytical forms we also represent the core charge density just as one single Gaussian. 
The amplitude and width of this Gaussian core charge distribution are then optimized by a fitting procedure in the same way as the other parameters of the pseudopotential. 

\section{Methodology}

The procedure for constructing the NLCC pseudopotentials is very similar to the one used for the construction of the GTH and HGH pseudopotential.
In contrast to the original GTH and HGH pseudopotentials which were fitted to a single atomic configuration, the new NLCC pseudopotentials are fitted not only to the ground state but also to several excited and ionized electronic configurations where half an electron is added or removed possibly to or from different valence orbitals.
The atoms are always considered to be spherically symmetric, but some of the configurations used for the fit have a spin polarization. 
The parameters of the dual space Gaussian ansatz~\cite{Goedecker96} are fitted such that: 
\begin{itemize}
\item The occupied and first few unoccupied valence eigenvalues of the all-electron and pseudo atom match for all configurations used in the fit.
\item The charge inside the inert region of the pseudo atom matches the all-electron valence charge in the same region for all the orbitals for which the eigenvalues are matched for all configurations used in the fit. This means that the pseudopotential is norm conserving for all the configurations used in the fit.
\item A high precision of $10^{-6}$ a.u. is achieved for valence eigenvalues and charge integrals of the non-polarized ground state, whereas only a moderate precision of $10^{-4}$ a.u. is enforced for all other orbitals and configurations considered.
\item The total energy differences of the all-electron atom are reproduced for all configurations used in the fit.
\item The spin polarization energies of the all-electron atom are reproduced for all spin polarized configurations used in the fit.
\end{itemize}
Since the considered quantities are fitted for several configurations atomic transferability is already built in to these new pseudopotentials by construction. 

The core charge is represented by a single Gaussian which is optimal for numerical efficiency.
It is initialized such that it approximates well the physical core charge density and it is held fixed during an initial stage of the fit. Then both the amplitude and the width of the Gaussian are released and considered as fitting parameters. As a consequence the total amount of core charge and the width can differ from the physical value. 


The parameters of the core charge constitute thus a small set of only two additional degrees of freedom. Yet this allows to optimally reproduce atomic polarization energies without degrading the transferability and accuracy of other atomic properties. It was found that the inclusion of a more complicated core charge is not beneficial. 
Furthermore, it should be emphasized that the novel NLCC pseudopotentials are not harder than their HGH counterparts. The smoothness of the core charge seems to play an important role for the fact that the hardness is not affected, and roughly the same grid spacings or energy cutoffs can be employed as for conventional HGH pseudopotentials. 

In particular, pseudopotentials with NLCC were generated for boron, carbon, nitrogen, oxygen, fluorine, aluminium, silicon, phosphorous, sulfur and chlorine.
Very weak spin dependences are expected for the rare gasses, and for all remaining chemical elements up to the third row, NLCC are found to be unnecessary, as HGH pseudopotentials are available that either include semicores (sodium and magnesium) or leave no core states at all (hydrogen, lithium and beryllium). For the special case of hydrogen, it was found that the multi configuration fit gave slightly improved results even though obviously no core charge was added.
Since the focus of this paper is on systems made out of light elements, no relativistic effects such as spin-orbit coupling were included in the pseudopotentials. 

\section{Results}

To assess the accuracy of the new pseudopotentials extensive calculations were performed for different test sets. The accuracy of covalent bond formation energies was examined for the standard G2-1 test set~\cite{G2-1,G1,G1x,G2x}. 
For the assessment of the accuracy of non-bonded interactions the S22~\cite{S22} test set was used.
To check the performance for materials under high pressure we chose carbon, silicon, silicon carbide and boron nitride as test systems. 

All pseudopotential calculations were done with the BigDFT package~\cite{Genovese08}. The BigDFT code uses a systematic wavelet basis set which allows to obtain the exact density functional solution with arbitrarily small error bounds. The parameter were set such that an accuracy of at least $10^{-6}$ Hartree was obtained.
The LibXC library~\cite{libxc} is used within BigDFT for the evaluation of the exchange correlation functional. 
Semi-empirical van-der Waals corrections were added in BigDFT according to the DFT-D2\cite{GrimmeD2} and DFT-D3\cite{GrimmeD3} methods for the calculations of the S22 test set.

To obtain reliable all-electron reference values for the atomization energies of the G2-1 test set, we performed all-electron calculations with the NWChem software package~\cite{NWChem} using one of the largest available Gaussian type basis sets, namely an augmented correlation consistent polarized valence quintuple zeta Gaussian type basis set (aug-cc-pV5Z).
Care was taken to disable symmetry detection and to check for the lowest energy spin multiplicity. For the chemical elements Li, Be Na and Mg, the aug-cc-pV5Z set was not available, so the corresponding quadruple zeta set (aug-cc-pVQZ) was used to compute the atomization energies of $Li_2$, $LiF$, $BeH$, $Na_2$, $NaCl$, $MgH$ and $Mg2$. To obtain the atomization energies of the relaxed molecules, geometry optimizations were carried out using the very same basis set.

Atomic all-electron calculations of the spin polarization energies were done with our non-spherical atomic code, which expresses the wavefunctions as a product of spherical harmonics and radial functions. The radial function are given numerically on a logarithmic grid. The settings were chosen such that a precision of at least $10^{-8}$ Hartree can be obtained for the total energy. 
This required angular integration grids of 232 points and multipole representations up to $\ell=4$.
The atomic LSDA reference energies from the National centre of science and technology (NIST)\cite{NISTatomic} where reproduced within the given precision for all elements considered.

We calculated the atomization energies also with the three different sets of PAW~\cite{PAW} potentials available in VASP~\cite{Vasp}. Those PAW potentials are derived from the all-electron atomic Hamiltonian and aim at all-electron accuracy. 
In order to obtain the required high precision some parameters had to be set to tighter values than the default values. 
We had to use for the general accuracy (\textsc{prec = High Accurate} and \textsc{lasph = true}) to activate nonspherical gradient corrections inside the PAW spheres.
It was carefully checked that the calculations were converged with respect to the size of the periodic simulation cell. Furthermore, care was taken that the correct spin multiplicity and non-fractional occupations were produced. Hard PAW potentials were available for all required elements except for Li, Be, Na and Mg, for which semicore potentials were used instead. For comparison, all energies were recomputed with a set of default potentials.
The third set consists of soft potentials for the elements B, C, N, O and F and default potentials otherwise.
For the periodic solids, all-electron calculations have been performed using the full-potential linearized augmented plane wave (FLAPW) and augmented plane wave plus local orbitals (APW + lo) methods as implemented in the WIEN2k\cite{Wien2k} software package.
We used a reduced muffin-tin radii for all atomic sorts in order to avoid their overlap up to the highest studied pressures. The sphere radii were kept fixed throughout the whole set of lattice parameters to obtain the best possible error cancellation. 
Semicore states were treated as valence, because high compressions can lead to an overlap of their wavefunctions, which will give a contribution to the energy.
Inside the spheres, the partial waves were expanded up to \textsc{lmax = 10}. The number of plane waves was limited by a cutoff parameter \textsc{RMTKmax = 9.0} for all the compounds under
consideration. The charge density was Fourier expanded with \textsc{Gmax} = 14 a.u. 
For the majority of the systems we used a very dense k-points grid (15$\times$15$\times$15) to ensure total energy convergence.

All the calculations were done at zero electronic temperature, i.e. no Fermi smearing was used. 
Zero point energies were not included in any of our results.

\subsection{Atomization energies of the G2-1 test}

Atomization energies are frequently used to assess the quality of various exchange correlation functionals as well as other approximations used in electronic structure calculations. The Gaussian G2-1 test set~\cite{G2-1,G1,G1x,G2x} is a standard benchmark set of 55 molecules in this context. Since this test set does not contain molecules with the chemical elements B, Al and Mg we added the molecules BH, BH2, AlH, AlH2, Mg2 und MgH.
We used this augmented test set to compare our pseudopotential results with all-electron calculations.
Because of Hund's rule most isolated atoms are strongly spin polarized. When an atom is inserted into a molecule or solid, its spin polarization is typically strongly reduced. Since standard pseudopotentials are based on a non-spin polarized reference configuration they can typically better describe atoms in molecules or solids than isolated atoms themselves. Since the atomization energy is the difference between the total energy of the molecule and the sum of the total energies of its constituent isolated atoms, the largest contribution to the error in the atomization energy of a pseudopotential calculation comes actually from the atomic energies. 

\begin{figure}
\includegraphics[angle=0, width=\figwidth]{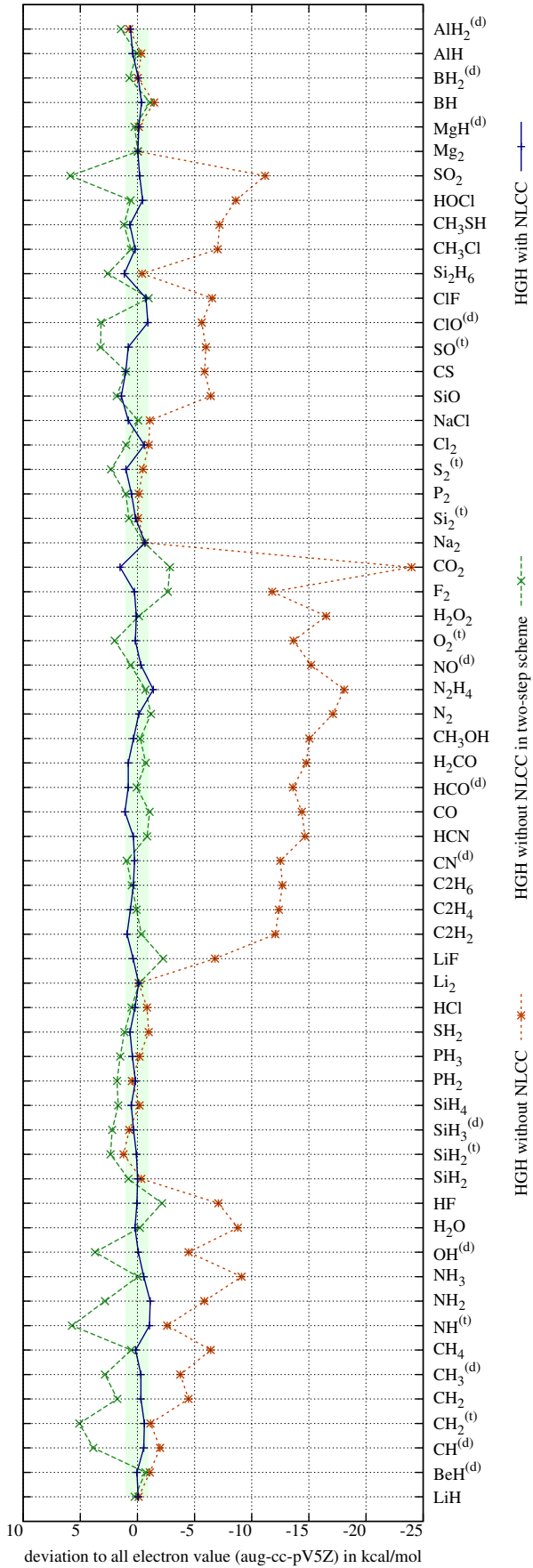}
\caption{ \label{HGH} Accuracy of PBE~\cite{PBE} atomization energies computed with HGH pseudopotentials. Explanations are in the text.}
\end{figure}

The atomization energies of the molecules in the G2-1 test set were first computed using conventional HGH pseudopotentials~\cite{Krack05} for the PBE exchange correlation functional. A comparison with all-electron data is shown in figure \ref{HGH}.
The spin multiplicity of systems with a net magnetic moment are indicated in brackets and omitted for closed shell systems. 
Deviations of $\pm 1$ kcal/mol are indicated with a (green) shading to relate the errors to the requirements for chemical accuracy. 

It is found that the direct computation of the electronic atomization energies with the conventional pseudopotentials leads to significant disagreement with the results obtained in an all-electron calculation. An rather high mean absolute deviation (MAD) of 6.83 kcal/mol to the electron reference values for all 55 molecules in the G2-1 set is found. 
However, the main contribution to the error in the atomization energies comes from the estimation of the energy of the isolated atoms. Therefore, the atomization energies can be improved significantly by a two step procedure where the atomization energies are calculated as a sum of two terms. The first term is the energy difference between the molecule and the sum of the total energies of isolated, spherical and non-spinpolarized atoms. It thus can be considered as the atomization energy with respect to a set of non-physical atoms. This energy difference is calculated with the HGH pseudopotentials and is fairly accurate since no strong spin polarizations are involved. 
The second term is the difference in total energy between the real, i.e non-spherical and spin polarized, atom and the previously defined non-physical atom. This second term is calculated with our all-electron program for non-spherically symmetric atoms and is therefore exact. 
Since the atomic spin polarization energies and energy terms for breaking the spherical symmetry are only a property of the atoms they can be considered as a set of atomic correction terms for the accurate calculation of atomization energies. The atomic correction terms for the chemical elements considered in this study are listed in Table \ref{tableAtomic}.

\begin{table}
\begin{tabular}{|c|c|c|c|c|c|c|c|c|c|c|c|c|}
\hline     H  &   Li &    B &    C &    N &    O &    F &  Na &  Al &   Si &    P &    S &  Cl \\
\hline   25.4 &  6.8 & 10.4 & 31.7 & 72.0 & 43.7 & 16.0 & 5.1 & 7.1 & 19.7 & 43.1 & 23.8 & 7.6 \\
\hline
\end{tabular} \\
\caption{ \label{tableAtomic}
Atomic correction terms in kcal/mol as used for the two step procedure.
} 
\end{table}

It has to be stressed that these atomic spin polarization energies drop out in most instances such as in the calculation of energy differences in a chemical reaction where only molecules are involved. 
Using this two step scheme, the errors in the atomization energies are decreased considerably to a MAD of 1.56 kcal/mol. 
Because of the cancellation effect, this is the accuracy that can be expected in the majority of energy differences calculated with the standard HGH type pseudopotentials. The above MAD value was obtained with the bond lengths and angles fixed as given on the computational chemistry comparison and benchmark database\cite{CCCBDB}. 
Using instead the equilibrium geometries of each method, i.e. the HGH pseudopotential and the all-electron calculation, the MAD value is slightly decreased to 1.52 kcal/mol.
This last value is actually more relevant in practice since atomization energies for an unknown system necessarily have be calculated with theoretically determined geometries. 


The new Gaussian type pseudopotential with a NLCC can however still considerably improve the accuracy without the need of using a two step procedure. The error of a direct computation of the atomization energies decreases to a MAD of only 0.52 kcal/mol. Using the equilibrium geometries obtained with the new NLCC pseudopotential the error drops again slightly to 0.51 kcal/mol. More important than this small improvement of the MAD is the fact that the result could be improved for the few molecules where the error was well above the average. 


\begin{figure}
\includegraphics[angle=0, width=\figwidth]{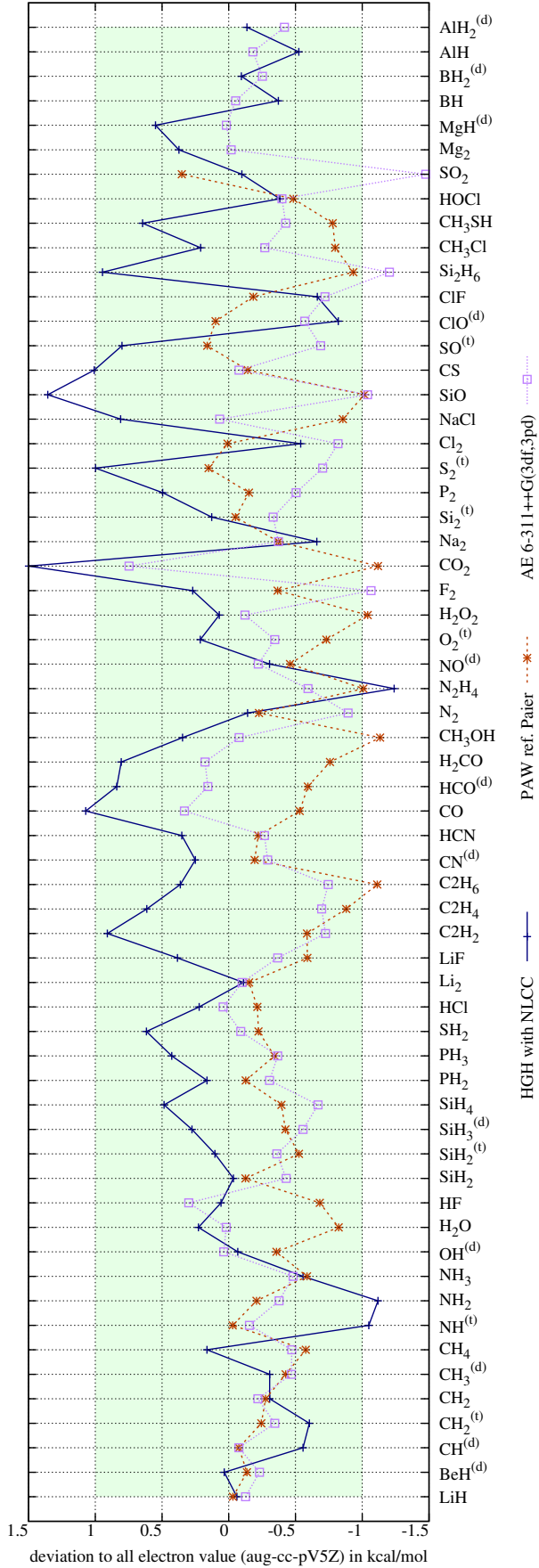}
\caption{ \label{comparegood} Comparison of PBE atomization energies from NLCC-HGH pseudopotentials with other methods. Explanations are in the text.}
\end{figure}

\begin{figure}
\includegraphics[angle=0, width=\figwidth]{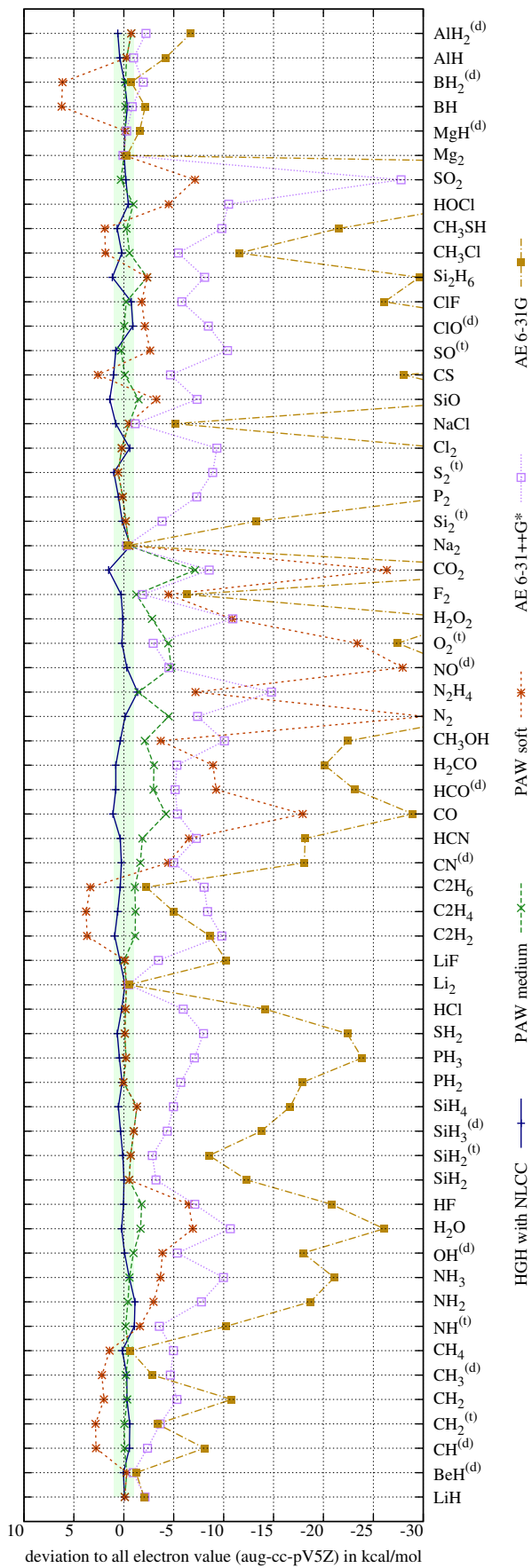}
\caption{ \label{comparebad} Comparison of PBE atomization energies from NLCC- HGH pseudopotentials with less accurate methods. Explanations are in the text.}
\end{figure}

In figure \ref{comparegood} the accuracy of the HGH pseudopotentials with NLCC is shown for relaxed molecular geometries and compared with the results of PAW calculations published by Paier et. al.\cite{Paier05}. In this work hard PAW potentials were used and we were able to reproduce their results. 
In essence, the absolute errors of the new NLCC pseudopotentials are comparable with those using hard PAW potentials. 
As shown in Figure \ref{comparebad} the accuracy however goes down significantly when one uses the default or even the soft PAW potentials of the VASP package~\cite{Vasp}. 
Furthermore, the same figures shows the discrepancies between all-electron results obtained in two large Gaussian basis sets while keeping the molecular geometries fixed. Even at this size the differences between the two basis sets are not negligible compared to the deviations to other methods and the accuracy of the pseudopotential method is indeed close to the discrepancies between different choices of all-electron reference values.
This is quite surprising given the fact that these simple chemical compounds show only straightforward covalent type bonding properties which are certainly easier to describe with a Gaussian basis set than other more complex bonding patterns. 
It has also to be stressed that the computational cost rises very steeply when one goes from a medium size basis set to these very large 
basis sets. This is in contrast to the wavelet method where a modest decrease of about 15 percent in the grid spacing $h$ results in an gain of a factor of ten in accuracy because of the high order convergence rate of $h^{14}$.

The accuracy problems of Gaussian basis sets become even more evident if one employs medium size or small standard basis sets in an all-electron calculation. 
The 6-31G, 6-31++G*, 6-31+G** and 6-311++G(3df,3pd) basis sets were employed to compare the relative accuracy of the pseudopotential method with the incompleteness of and disagreement between standard Gaussian basis of various sizes. 
Figure \ref{comparebad} clearly shows that the accuracy obtained with these basis set is considerably lower than the accuracy with the NLCC pseudopotentials or also with the standard HGH pseudopotential within the two step procedure described above.

A summary of the deviations in the atomization energies averaged over the molecules of the G2-1 test set is given for fixed and relaxed geometries in tables \ref{tableFixed} and \ref{tableOpt}, respectively. Indicated are the MAD, RMSD, mean signed deviation (MSD), maximum absolute deviation (maxAD) and minimum absolute deviation (minAD). 

The last row of table \ref{tableOpt} describes the change in the all-electron reference values when going from the fixed, experimental (CCCBDB) to relaxed geometries in the aug-cc-pV5Z basis set. This gain in energy upon geometry relaxation is significant compared to the assessed accuracy of the pseudopotential based methods, which are found to be very reliable for geometry optimizations. 

\begin{table}
\begin{tabular}{|l|r|r|r|r|r|}
\hline                   &   MAD   &  RMSD &    MSD &   maxAD & minAD \\
\hline          NLCC-HGH &  0.52   &  0.65 &  0.15  &   1.52  &  0.01 \\
\hline         HGH Krack &  6.82   &  9.13 & -6.74  &  23.98  &  0.05 \\
\hline          Two-step &  1.56   &  2.09 &  0.91  &   5.86  &  0.04 \\
\hline        PAW medium &  1.20   &  1.89 & -1.14  &   7.15  &  0.01 \\
\hline          PAW soft &  4.84   &  8.53 & -3.77  &  30.23  &  0.05 \\
\hline 6-311++G(3df,3pd) &  0.43   &  0.53 & -0.36  &   1.48  &  0.02 \\
\hline          6-31++G* &  6.53   &  7.76 & -6.53  &  27.78  &  0.27 \\
\hline             631-G & 22.13   & 31.16 &-22.13  & 151.88  &  0.37 \\
\hline
\end{tabular} \\
\caption{ \label{tableFixed}
Deviation measures of the electronic atomization energies in kcal/mol for the 55 molecules of the G2-1 test set compared to the all-electron result obtained in the aug-cc-pV5Z basis set. All geometries are fixed.
} 
\end{table}

\begin{table}
\begin{tabular}{|l|r|r|r|r|r|}
\hline              &   MAD   &  RMSD &    MSD &   maxAD & minAD \\
\hline         NLCC &  0.51   &  0.63 &   0.16 &   1.50  &  0.03 \\
\hline    HGH Krack &  6.85   &  9.13 &  -6.76 &  23.94  &  0.10 \\
\hline     Two-step &  1.52   &  2.05 &   0.88 &   5.73  &  0.01 \\
\hline    PAW Paier &  0.46   &  0.56 &  -0.43 &   1.13  &  0.01 \\
\hline
\hline   all-electron geopt  &  0.29   &  0.70 & -0.29  &   4.21  &  0.00 \\
\hline
\end{tabular} \\
\caption{ \label{tableOpt}
Deviation measures of the electronic atomization energies in kcal/mol for the 55 molecules of the G2-1 test set, where all molecular geometries are optimized for each method considered. For comparison, PAW data are extracted from work of Paier et. al\cite{Paier05}. The last row gives the change of the all-electron energy upon geometry relaxation. 
} 
\end{table}

\subsection{Accuracy of the equilibrium geometries}

\begin{figure}
\includegraphics[angle=0, width=\figwidth]{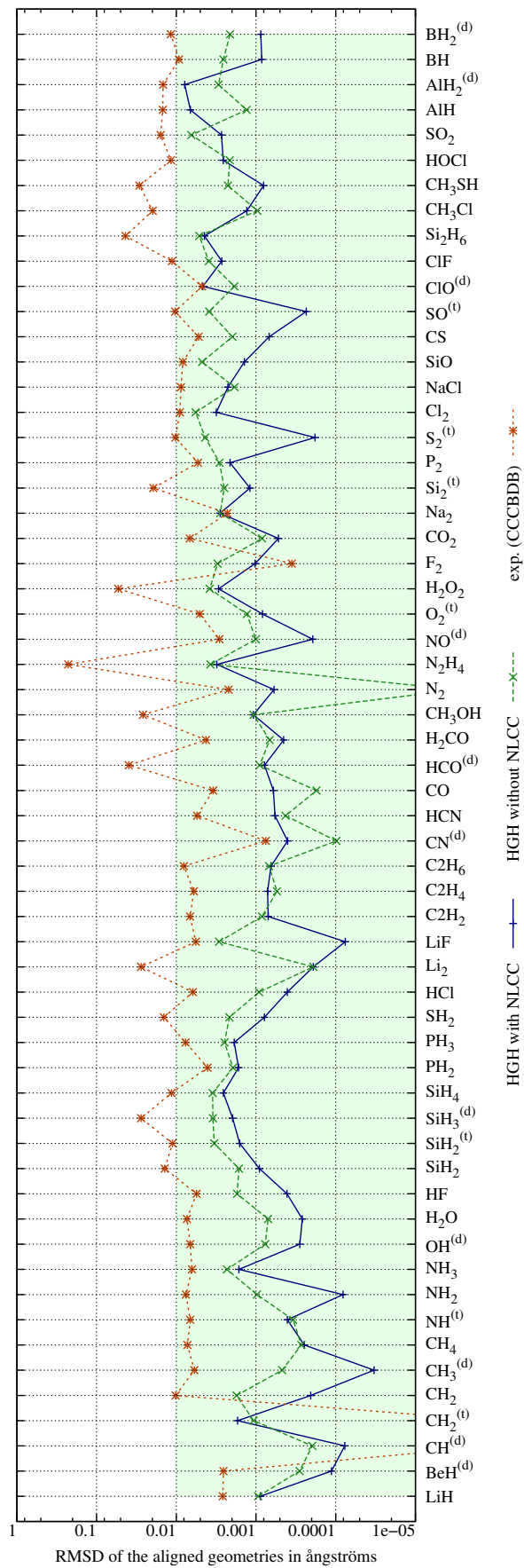}
\caption{ \label{rmsd} RMSD values of the experimental and relaxed geometries with respect to those relaxed with all-electron calculations in the aug-cc-PV5Z set.}
\end{figure}

In order to compare the accuracy of the equilibrium geometries of the pseudopotential and all-electron calculations, the optimized geometry of each molecule is aligned with its all-electron counterpart, such that the RMSD is minimized~\cite{Kabsch76}. 
The resulting RMSD values are shown in figure \ref{rmsd}. It is observed that conventional HGH pseudopotentials yield already very good agreement with the all-electron data. Nevertheless, the inclusion of NLCC leads to a systematic improvement of the equilibrium geometries.

It can be noted that in a previously mentioned work~\cite{Paier05}, a similar test was carried out for the bond lengths of some dimers in order to verify the accuracy of the PAW method. It is found that our pseudopotential approach yields geometry data of at least the same or even better accuracy, and that the high precision is maintained when moving to more complicated geometries.

\subsection{Evaluation of pressure of extended systems}

Next we benchmark pseudopotential (PSP) calculations for extended systems. A few crystalline systems made of light elements (diamond carbon, Silicon Carbide, Bulk Silicon and Boron Nitride) were selected and the pressure at a given lattice parameter was then compared between different approaches.

The details on how the stress energy tensor can be calculated in GGA for NLCC terms are given in the appendix.
Figure \ref{Pressfig} shows the difference between the LAPW results and the PSP results at the same lattice parameter. 
In addition we also show the results for the hard PAW potentials. 
In order to show the relative accuracy of the pressure we specify the lattice constant along the x axis of the figure in terms of the pressure. At the highest pressures the lattice constant is reduced by about 10 percent compared to the lattice constant with zero pressure. 
With NLCC PSP, an excellent relative accuracy of about $10^{-3}$ is found.
In this case, it can be noticed that the inclusion of a NLCC term improves further the results even though the systems under pressure are not spin-polarized.
Results of similar quality can be obtained within the hard PAW scheme described above.

\begin{figure}
\includegraphics[angle=-90, width=\figwidth]{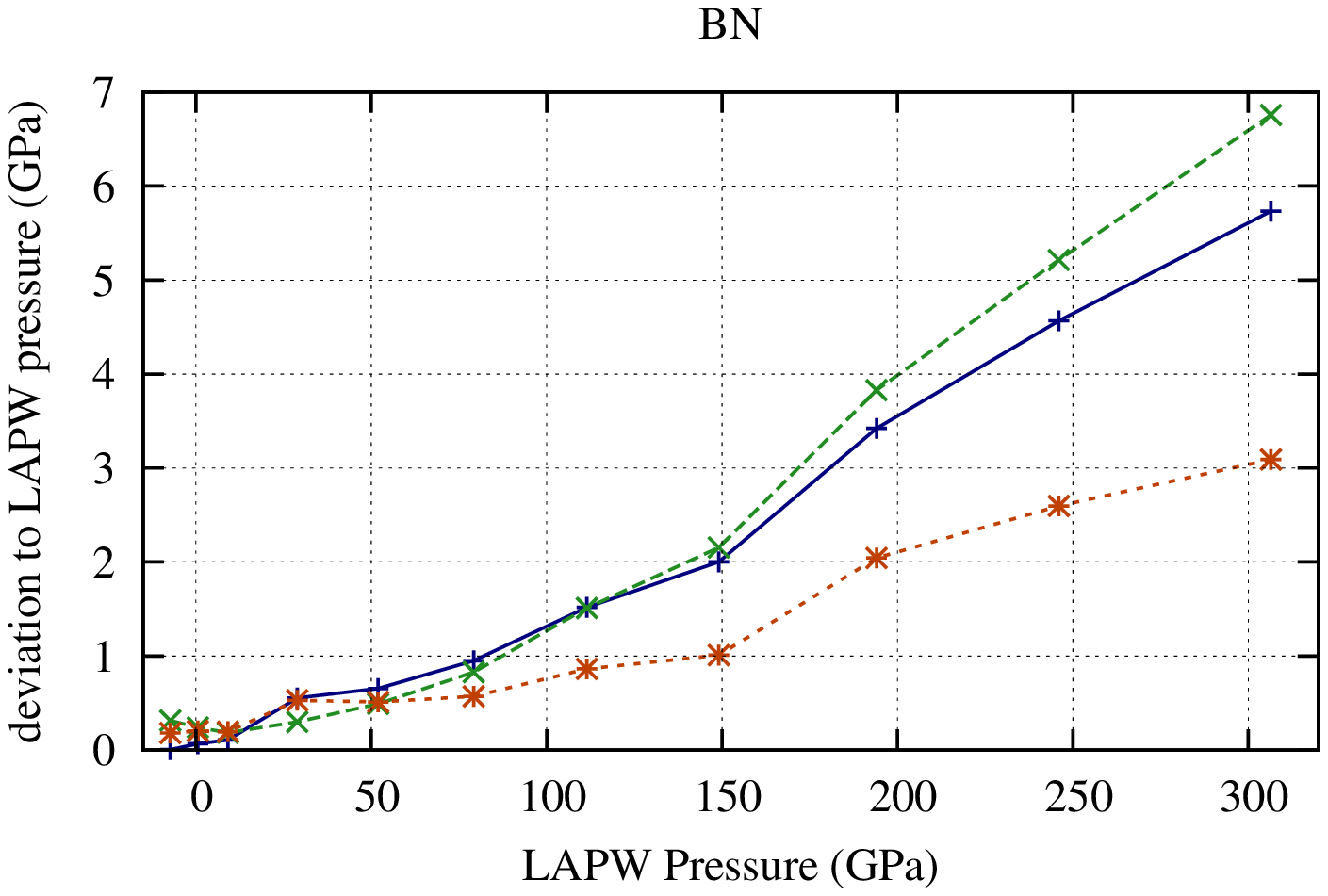}\\
\includegraphics[angle=-90, width=\figwidth]{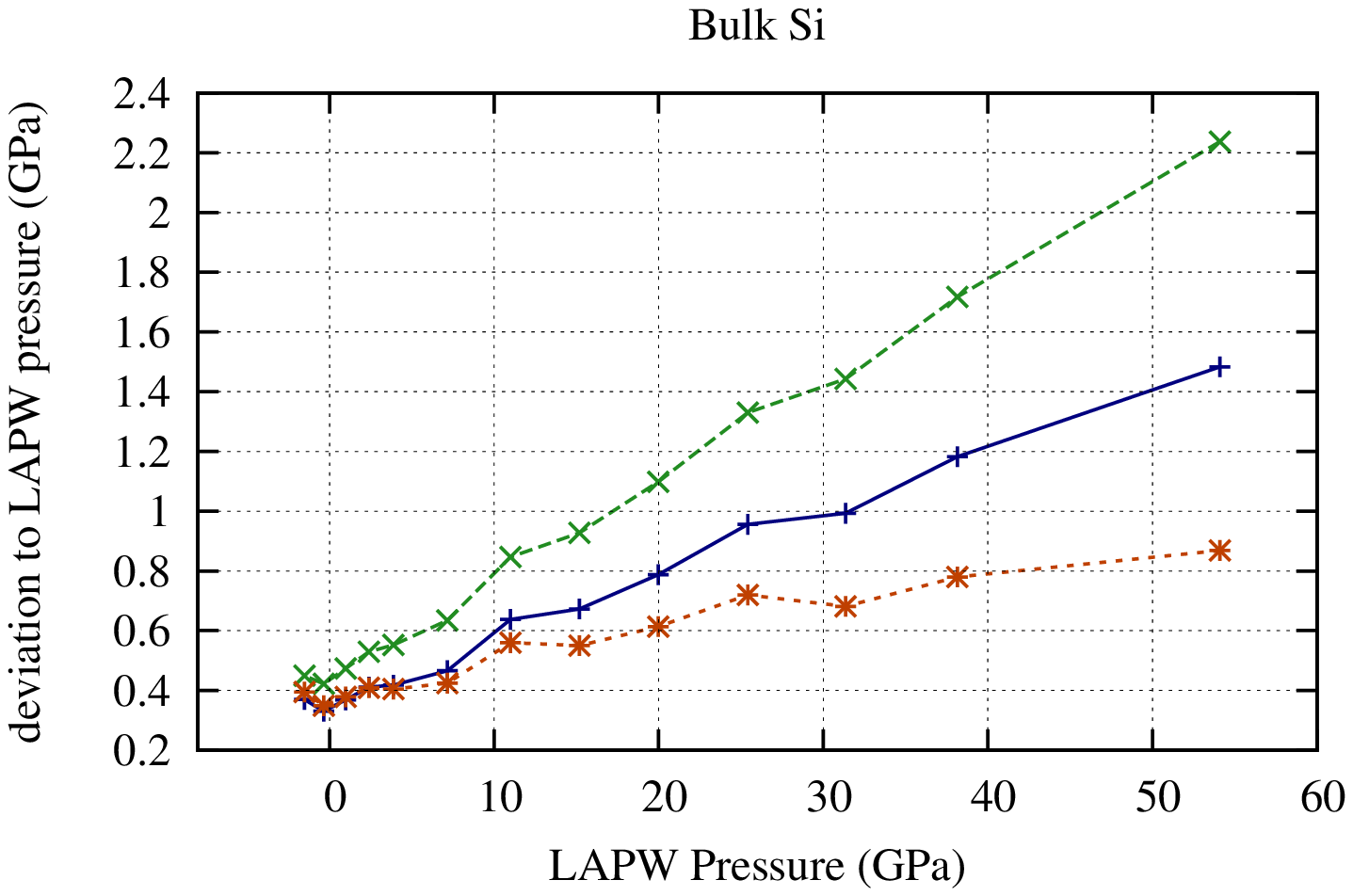}\\
\includegraphics[angle=-90, width=\figwidth]{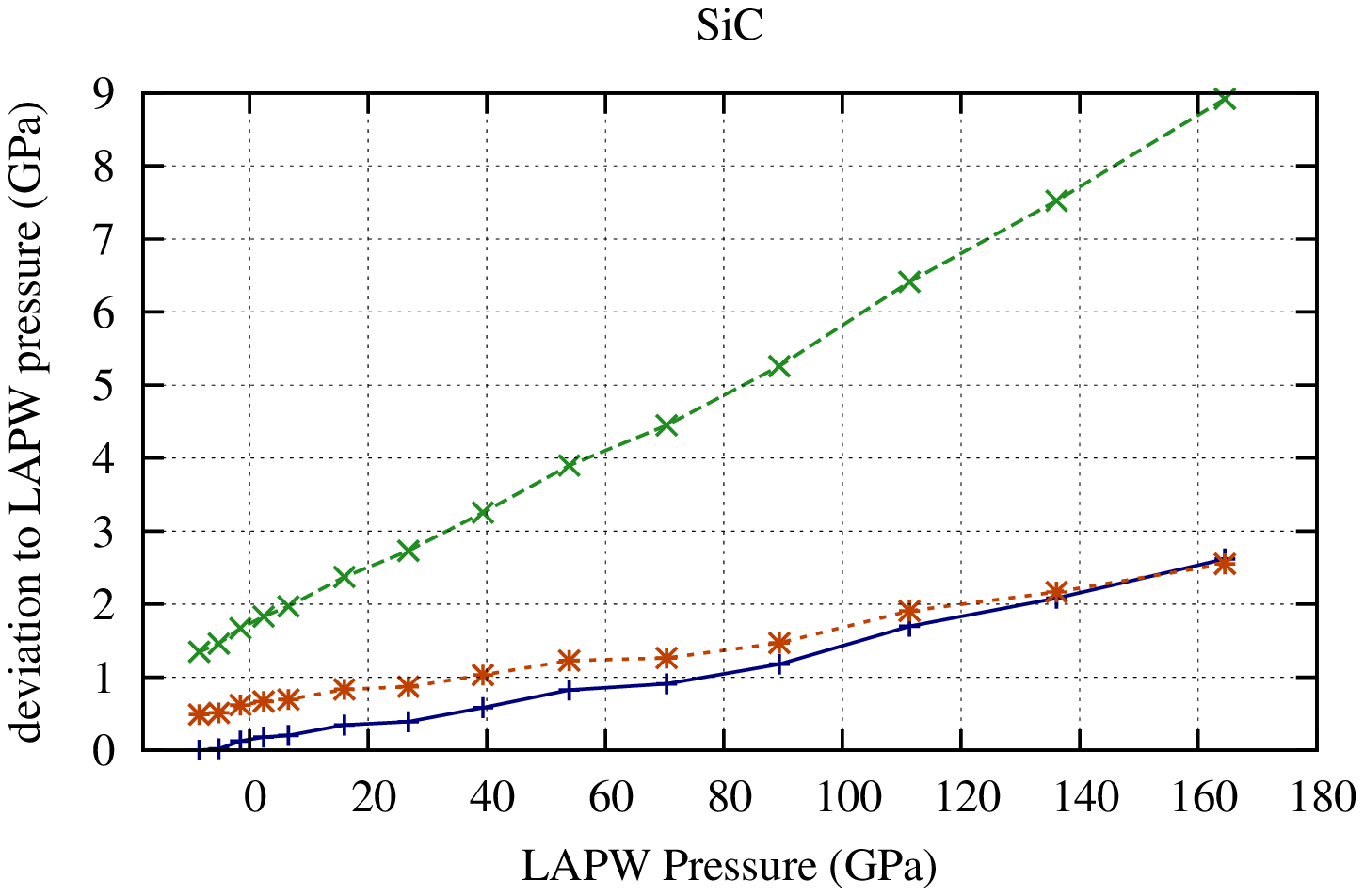}\\
\includegraphics[angle=-90, width=\figwidth]{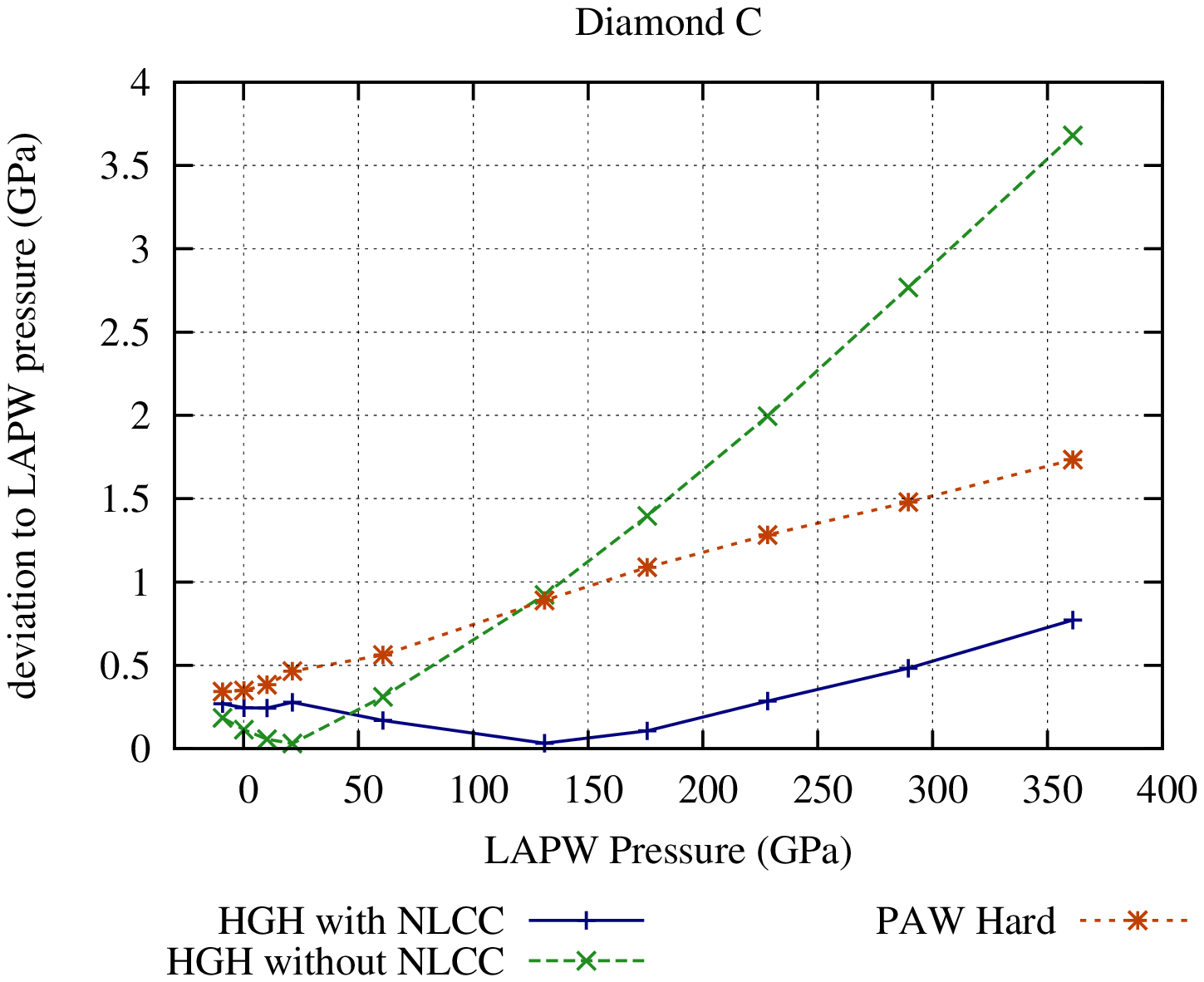}
\caption{ \label{Pressfig} Comparison of pressures.}
\end{figure}

\subsection{Dispersion-corrected functionals}

Long range van der Waals interactions are missing in all standard LDA and GGA density functionals. 
Adding semiempirical classical van der Waals interactions has however recently been demonstrated to give a rather accurate description of weakly bonded systems and is now frequently used. We will therefore examine the accuracy of the semiempirical models in the context of our pseudopotential calculations with a systematic wavelet basis set. 

\begin{figure}
\includegraphics[angle=0, width=\figwidth]{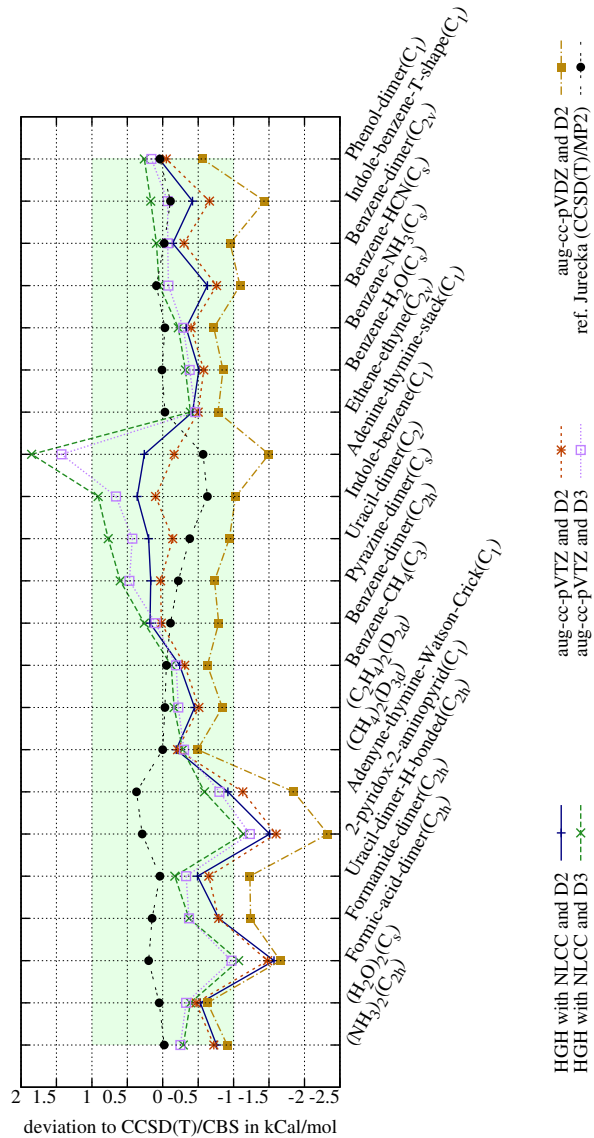}
\caption{ \label{S22fig} Comparison of S22 test set results between PSP and all-electron calculations within PBE XC functional.}
\end{figure}

In BigDFT, we implemented two semiempirical models to correct dispersion energies and energy gradients DFT-D2\cite{GrimmeD2} and DFT-D3\cite{GrimmeD3}. The parameters of these models were separately fitted for each exchange correlation functional based on thermochemical data for weakly interacting systems. Since BigDFT uses a wavelet basis and pseudopotentials Figure \ref{S22fig} and Table \ref{S22tab} show the comparison of interaction energies of the benchmark database S22\cite{S22}, with a reference calculation using Coupled Cluster CCSD(T) in the complete basis limit (CBS)\cite{Sherril}.
The inclusion of dispersion correction D2 into BigDFT clearly improves the description of weak interactions within PBE, even though the S22 data set was not used as the fitting data set. The root-mean-square-deviation (RMSD) between the CCSD(T) reference values and the NLCC-DFT interaction energies is 0.58 kcal/mol. 
The absolute maximum difference corresponds to acetic acid dimer (COOH)$_2$, where the overestimation is 1.57 kcal/mol, that means an 8\% of the total interaction energy. The largest relative error of 35 \% is found for the methane dimer whose interaction energy is only 0.2 kcal/mol. 
The errors for these systems are comparable to those that are obtained when PBE-D2 and PBE-D3 are used with a large basis set (aug-cc-pVDZ and aug-cc-pVTZ).
On average the PBE-D2 scheme performs better with BigDFT than with any Gaussian basis set, while the PBE-D3/BigDFT results are comparable to the results obtained with PBE-D3/aug-cc-pVTZ electron calculations. 

\begin{table}
\begin{tabular}{|l|r|r|r|r|r|}
\hline               &   MAD  & RMSD &    MSD &  maxAD & minAD \\
\hline NLCC-no corr. &  2.59  & 3.61 &   2.59 & 10.08  &  0.05 \\
\hline HGH Krack-no corr. &
                       2.64   & 3.66 &   2.64 & 10.18  &  0.01 \\
\hline NLCC-D2       &  0.51  & 0.64 &  -0.39 &  1.57  &  0.05 \\
\hline HGH Krack-D2  &  0.50  & 0.58 &  -0.34 &  1.25  &  0.13 \\
\hline NLCC-D3       &  0.48  & 0.64 &  -0.03 &  1.14  &  0.05 \\
\hline HGH Krack-D3  &  0.47  & 0.64 &   0.02 &  1.95  &  0.03 \\
\hline aug-cc-pVDZ-D2&  1.05  & 1.15 &  -1.05 &  2.33  &  0.49 \\
\hline aug-cc-pVTZ-D2&  0.53  & 0.68 &  -0.51 &  1.60  &  0.02 \\
\hline aug-cc-pVTZ-D3&  0.44  & 0.57 &  -0.14 &  1.43  &  0.07 \\
\hline
\end{tabular} \\
\caption{ \label{S22tab}
Deviation measures in kcal/mol for the S22 test set with respect to CCSD(T) calculations. For the PBE XC functional, PSP and all-electron calculations are compared including semiempirical dispersion corrections (D2 and D3).
} 
\end{table}

\section{Discussion and conclusions}

We have shown that our new NLCC PSP's give very high accuracy for a wide range of applications. In particular they give atomization energies with chemical accuracy compared to all-electron calculations for the G2-1 test set. This accuracy can easily be obtained with a systematic basis set such as wavelets where one has to change only a single parameter to obtain arbitrarily high accuracy. Obtaining such a high accuracy with Gaussian basis sets requires using the largest available basis sets and is therefore frequently not feasible in practice. 
Contrary to a widespread belief, PAW calculations do not necessarily give all-electron accuracy. Soft PAW potentials can actually lead to appreciable errors. Well constructed hard PAW potentials on the other hand give very high accuracy and are together with our new norm-conserving pseudopotentials in practice the only feasible way to highest quality results for large systems. 


\section{Appendix: NLCC HGH pseudopotentials in Kohn-sham DFT formalism} 

The PSP format is based on HGH-Krack form~\cite{Krack05}:
\begin{equation}
\hat V_{PSP} = \hat V_{loc} + \hat V_\text{nl}\,,
\end{equation}
where the first part is a local potential
\begin{multline}\label{GTHloc}
V_{loc}(r) = -\frac{Z_{ion}}{r} \mathrm{erf} \left( \frac{r}{\sqrt{2} r_{loc}} \right) \\
           +  \exp\left( -\frac{r^2}{~2~ r_{loc}^2} \right)  
\left( \sum_{k=1}^{n\leq 4} c_k \left(\frac{r}{r_{loc}}\right)^{2k-2} \right)
\end{multline}
and the second part the non-local term which is separated into different channels 
$\hat V_\text{nl}=\sum_\ell V_{\ell}({\bf r,r'})$,
each one defined in terms of separable projectors

\begin{align}\label{GTHsep}
V_{\ell}({\bf r,r'}) &= \sum_{i,j=1}^{n\leq2} \sum_{m=-\ell}^{\ell}
 p_{i}^{\ell m}({\bf r}) ~
h_{ij}^{\ell} ~
p_{j}^{\ell m}({\bf r}')\\
p_{i}^{\ell m}({\bf r}) &= \frac{\sqrt{2} ~ r^{2\ell+i} 
e^{ \left(-\frac{1}{2}\left(\frac{r}{r_{\ell}}\right)^2 \right) }}{r_{\ell}^{\ell+(4i-1)/2} \sqrt{ \Gamma
\left( \ell + \frac{4i-1}{2}  \right) } } Y_{\ell m}(\theta,\phi) \;. \label{GTHproj}
\end{align}
The core charge $\rho_c$ of the new PSP's is given by 
\begin{equation} \label{GTHcore}
\rho_c(r) = c_\text{core} \frac{Z-Z_{ion}}
{\left( \sqrt{2 \pi} r_\text{core} \right)^3 } e^{-\frac{r^2}{2 r_\text{core}^2}}\;.
\end{equation}

The pseudopotentials parameters according to equations \eqref{GTHloc} to \eqref{GTHcore} are given in table \ref{tablePSP}.

\begin{table}
\begin{tabular}{|l| r r r |r r r|}
\hline
 H   &     1       &    1       &              &$Z        $&$  Z_{ion}$  &            \\
     &     0.20000 &   -4.07312 &    0.68070   &$r_{loc}  $&$  c_1      $&$ c_2     $ \\
\hline                                        
 B   &     5       &    3       &              &$Z        $&$  Z_{ion}$  &            \\
     &     0.43250 &   -4.26853 &    0.59951   &$r_{loc}  $&$  c_1      $&$ c_2     $ \\
     &     0.37147 &    6.30164 &              &$r_s      $&$  h^{s}_{11} $&$       $ \\
     &     0.33352 &    0.43364 &              &$r_{core} $&$  c_{core} $&$         $ \\
\hline                                        
 C   &     6       &    4       &              &$Z        $&$  Z_{ion}$  &            \\
     &     0.31479 &   -6.92377 &    0.96360   &$r_{loc}  $&$  c_1      $&$ c_2     $ \\
     &     0.30228 &    9.57595 &              &$r_s      $&$  h^{s}_{11}  $&$      $ \\
     &     0.36878 &   -0.00996 &              &$r_p      $&$  h^{p}_{11}  $&$      $ \\
     &     0.27440 &    0.76008 &              &$r_{core} $&$  c_{core} $&$         $ \\
\hline                                        
 N   &     7       &    5       &              &$Z        $&$  Z_{ion}$  &            \\
     &     0.24180 &  -10.04328 &    1.39719   &$r_{loc}  $&$  c_1      $&$ c_2     $ \\
     &     0.25697 &   12.96802 &              &$r_s      $&$  h^{s}_{11}  $&$      $ \\
     &     0.15686 &   -0.73453 &              &$r_p      $&$  h^{p}_{11}  $&$      $ \\
     &     0.24612 &    0.66037 &              &$r_{core} $&$  c_{core} $&$         $ \\
\hline                                        
 O   &     8       &    6       &              &$Z        $&$  Z_{ion}$  &            \\
     &     0.26100 &  -14.15181 &    1.97830   &$r_{loc}  $&$  c_1      $&$ c_2     $ \\
     &     0.22308 &   18.37181 &              &$r_s      $&$  h^{s}_{11}  $&$      $ \\
     &     0.26844 &    0.10004 &              &$r_p      $&$  h^{p}_{11}  $&$      $ \\
     &     0.25234 &    0.44314 &              &$r_{core} $&$  c_{core} $&$         $ \\
\hline                                        
 F   &     9   &        7       &              &$Z        $&$  Z_{ion}$  &            \\
     &     0.20610 &  -19.86716 &    2.79309   &$r_{loc}  $&$  c_1      $&$ c_2     $ \\
     &     0.19518 &   23.47047 &              &$r_s      $&$  h^{s}_{11}  $&$      $ \\
     &     0.17154 &    0.61254 &              &$r_{core} $&$  c_{core} $&$         $ \\
\hline                                        
 Al  &     13      &    3       &              &$Z        $&$  Z_{ion}$  &            \\
     &     0.35000 &   -1.20404 &   -2.14849   &$r_{loc}  $&$  c_1      $&$ c_2     $ \\
     &     0.46846 &    2.69262 &    0.00000   &$r_s      $&$  h^{s}_{11}  $&$ h^{s}_{21} $ \\
     &             &            &    2.15425   &$         $&$              $&$ h^{s}_{22} $ \\
     &     0.54697 &    2.13804 &              &$r_p      $&$  h^{p}_{11}  $&$      $ \\
     &     0.48775 &    0.38780 &              &$r_{core} $&$  c_{core} $&$         $ \\
\hline                                        
 Si  &     14      &    4       &              &$Z        $&$  Z_{ion}$  &            \\
     &     0.33000 &   -0.07846 &   -0.79378   &$r_{loc}  $&$  c_1      $&$ c_2     $ \\
     &     0.42179 &    2.87392 &    0.02559   &$r_s      $&$  h^{s}_{11}  $&$ h^{s}_{21} $ \\
     &             &    0.00000 &    2.59458   &$         $&$              $&$ h^{s}_{22} $ \\              
     &     0.48800 &    2.47963 &              &$r_p      $&$  h^{p}_{11}  $&$         $ \\
     &     0.44279 &    0.41540 &              &$r_{core} $&$  c_{core} $&$         $ \\
\hline                                        
 P   &     15      &    5       &              &$Z        $&$  Z_{ion}$  &            \\
     &     0.34000 &   -1.62258 &   -0.72412   &$r_{loc}  $&$  c_1      $&$ c_2     $ \\
     &     0.38209 &    3.47754 &   -0.01267   &$r_s      $&$  h^{s}_{11}  $&$ h^{s}_{21} $ \\
     &             &            &    3.47461   &$         $&$              $&$ h^{s}_{22} $ \\  
     &     0.43411 &    3.37859 &              &$r_p      $&$  h^{p}_{11}  $&$         $ \\
     &     0.39868 &    0.45667 &              &$r_{core} $&$  c_{core} $&$         $ \\
\hline                                        
 S   &     16      &    6       &              &$Z        $&$  Z_{ion}$  &            \\
     &     0.33000 &    1.49043 &   -0.73314   &$r_{loc}  $&$  c_1      $&$ c_2     $ \\
     &     0.37046 &    6.18605 &    0.00000   &$r_s      $&$  h^{s}_{11}  $&$ h^{s}_{21} $ \\
     &             &            &    2.57761   &$         $&$              $&$ h^{s}_{22} $ \\  
     &     0.39772 &    3.89113 &              &$r_p      $&$  h^{p}_{11}  $&$         $ \\
     &     0.38622 &    0.57500 &              &$r_{core} $&$  c_{core} $&$         $ \\
\hline                                        
 Cl  &     17      &    7       &              &$Z        $&$  Z_{ion}$  &            \\
     &     0.32000 &   -0.27448 &              &$r_{loc}  $&$  c_1      $&$         $ \\
     &     0.32659 &    4.20336 &    0.00000   &$r_s      $&$  h^{s}_{11}  $&$ h^{s}_{21} $ \\
     &             &            &    4.55652   &$         $&$              $&$ h^{s}_{22} $ \\  
     &     0.36757 &    4.22908 &              &$r_p      $&$  h^{p}_{11}  $&$         $ \\
     &     0.42148 &    0.29324 &              &$r_{core} $&$  c_{core} $&$         $ \\
\hline
\end{tabular} \\
\caption{ \label{tablePSP}
Pseudopotential parameters of HGH potentials with NLCC. The ionic charge $Z_{ion}$, local radius $r_{loc}$ and coefficients $C_k$ define the local part \eqref{GTHloc}, while the separable part \eqref{GTHsep} is determined by the localization radii $r_{\ell}$ and coefficients $h_{ij}^{\ell}$. The parameters for the core charge \eqref{GTHcore} are $c_{core}$ and $r_{core}$.
} 
\end{table}

The core density is then used in the Kohn-Sham total energy expression as follows:
\begin{multline}\label{EKS}
E_\text{KS}= \sum_i \langle \psi_i | \bigl\{ -\frac{1}{2} \nabla^2 +
V_H[\rho] + V_{xc}[\rho+\rho_c] + V_{PSP} \bigr\}| \psi_i \rangle \\- E_H[\rho] + E_{xc}[\rho+\rho_c] -\int
\dd \r \rho(\r) V_{xc}[\rho+\rho_c](\r)\;,
\end{multline}
where $E_{xc}$ and $V_{xc}[n]=\frac{\delta E_{xc}[n]}{\delta n}$ are the XC energy and potential respectively, $V_H$ is the Hartree potential and the $\psi_i$'s are KS wavefunctions, whose summed squares give 
the valence density $\rho=\sum_i |\psi_i|^2$.
Eq.\ref{EKS} ensures Hellmann-Feynman condition at self-consistency, $\frac{\delta E_\text{KS}}{\delta \rho} =0$.




The contribution to the stress tensor $T_{\alpha \beta}^{xc}$ coming from the XC term with NLCC can be shown to be given by 
\begin{multline}
\Omega \,T_{\alpha \beta}^{xc} = \delta_{\alpha \beta} E_{xc}[\rho+\rho_c] -  \delta_{\alpha \beta} \int \dd \r
V_{xc}[\rho+\rho_c](\r) \rho(\r)\\   + \int \dd \r V_{xc}[\rho+\rho_c](\r) r_{(\alpha} \partial_{\beta)}\rho_c(\r) \\
- \int \dd \r \left(n(\r) \varepsilon[n]^{(2)}(\r) \frac{\partial_{(\alpha} n(\r)}{|\nabla n(\r)|}\right)
\partial_{\beta)} n(\r) \Biggr|_{n=\rho+\rho_c}\;,
\end{multline}
where $\Omega$ is the supercell volume and $\varepsilon[n]^{(2)}=\partial \varepsilon[n] / \partial (|\nabla n|)$.
The formula shows that the gradient of $\rho_c$ is needed to evaluate $T_{\alpha \beta}^{xc}$, even for a LDA computation.
A detailed derivation of DFT of stress (without NLCC) was shown by Dal Corso and Cresta~\cite{DalCorso94}. 

\section{Acknowledgements}

We acknowledge support from the Swiss National Science Foundation (SNF). Computer calculations were also performed at the Swiss National Supercomputing Center (CSCS) in Lugano. Partially, this research used resources of the Argonne Leadership Computing Facility at Argonne National Laboratory, which is supported by the Office of Science of the U.S. Department of Energy under contract DE-AC02-06CH11357.

\bibliographystyle{unsrt}

\end{document}